\documentclass[prl,twocolumn]{revtex4}
\usepackage{graphicx}
\usepackage{hyperref}
\usepackage{ulem}
\usepackage[section]{placeins}
\usepackage{amsmath}
\usepackage{amssymb}
\usepackage{dsfont}
\usepackage{bm}
\usepackage{tikz}

\newcommand{\ket}[1]{|#1\rangle}             % Ket Dirac's notation %
\newcommand{\be}{\begin{equation}}
\newcommand{\bea}{\begin{eqnarray}}
\newcommand{\ee}{\end{equation}}
\newcommand{\eea}{\end{eqnarray}}
\newcommand{\ben}{\begin{equation*}}
\newcommand{\bean}{\begin{eqnarray*}}
\newcommand{\een}{\end{equation*}}
\newcommand{\eean}{\end{eqnarray*}}
\newcommand{\ba}{\begin{align}}
\newcommand{\ea}{\end{align}}
\newcommand{\ban}{\begin{align*}}
\newcommand{\ean}{\end{align*}}

\usepackage{mathtools} % Enhances appearance of documents containing a lot of mathematics

\usepackage{xcolor} % Provides support for text colours

\usepackage{verbatim}

\usepackage{float} % Im­proves the in­ter­face for defin­ing float­ing ob­jects such as fig­ures and ta­bles.

\begin{document}

\title{Contrast inverted ghost imaging with non-interacting photons}

\author{Nicholas Bornman,$^{1,*}$ Megan Agnew,$^{2,*}$ Feng Zhu,$^{2}$ Adam Vall\'es,$^{1}$, Andrew Forbes,$^{1}$ Jonathan Leach$^{2}$}
\affiliation{$^1$School of Physics, University of Witwatersrand, Johannesburg 2050, South Africa}
\affiliation{$^2$IPaQS, SUPA, Heriot-Watt University, Edinburgh EH14 4AS, United Kingdom}
\affiliation{$^*$These authors contributed equally}
\date{\today}

\begin{abstract}
Ghost imaging is the remarkable process where an image can be formed from photons that have not ``seen'' the object. Traditionally this phenomenon has required initially correlated but spatially separated photons, e.g., one to interact with the object and the other to form the image, and has been observed in many physical situations, spanning both the quantum and classical regimes.  To date, all instances of ghost imaging record an image with the same contrast as the object, i.e., where the object is bright, the image is also bright, and vice versa.  Here we observe ghost imaging in a new system - a system based on photons that have never interacted.  We utilise entanglement swapping between independent pairs of spatially entangled photons to establish position correlations between two initially independent photons.   As a consequence of an anti-symmetric projection in the entanglement swapping process, the recorded image is the contrast reversed version of the object, i.e., where the object is bright, the image is dark, and vice versa.  The results highlight the importance of state projection in this ghost imaging process and provides a pathway to teleporting images across a quantum network.  
\end{abstract}
\maketitle

\begin{figure}
%\centering
\includegraphics[]{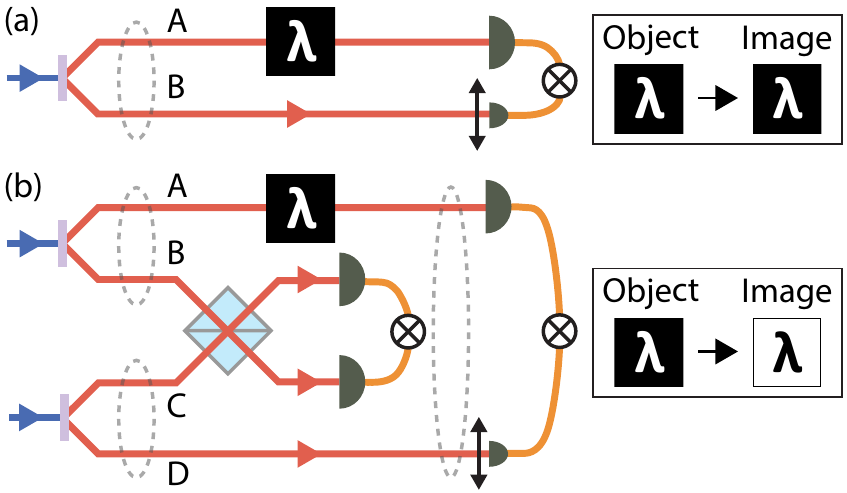}
\caption{\textbf{Two- and four-photon ghost imaging, using entangled photon pairs.} (a) Two-photon ghost imaging using photon pairs generated by parametric down-conversion. (b) Four-photon ghost imaging showing contrast inversion.  In the four-photon case, two pairs of photons are generated at two independent parametric down-conversion processes.  The state of the photons A and D become entangled when photons B and C undergo a Bell state measurement.  Contrast reversal of the object is observed when photons B and C are projected onto the anti-symmetric state.} 
\label{fig:4way}
\end{figure}

The term ``ghost imaging'' arose only in 1995 in the context of studying EPR correlations in position and momentum \cite{pittman1995optical}.  It was noted that the position correlations of an entangled photon pair generated by means of spontaneous parametric down-conversion (SPDC) could be used in an imaging experiment, as shown in Fig.~\ref{fig:4way}(a).  In a conventional ghost imaging experiment, the photon in the object arm is detected with a ``bucket'' detector with no spatial resolution so that the object information is erased, while the photon in the other arm, which has never interacted with the object, is collected with a spatially resolved detector (camera or scanning system).  Consequently, each photon has no information of the object, yet when the photons are measured in coincidence, the spatial correlations allow image reconstruction. Since this seminal work, many manifestations of the above have been realised, including the use of thermal light \cite{bennink2002two, bennink2004quantum, valencia2005two}, momentum correlated ghost imaging \cite{howell2004realization}, spiral ghost imaging with orbital angular momentum \cite{jack2009holographic, chen2014quantum}, time domain ghost imaging \cite{ryczkowski2016ghost}, computational and compressive ghost imaging \cite{shapiro2008computational, katz2009compressive}, and the use of non-degenerate SPDC for dual wavelength ghost imaging \cite{chan2009two, karmakar2010two}.

In all such cases the required correlations (in position or momentum) are established at the source by the momentum conservation law in the SPDC process, \cite{walborn2010}, at a beamsplitter with thermal light, or through knowledge of the illumination pattern in the case of computational ghost imaging. Implicit in the above is that it is not possible to demonstrate ghost imaging if the two photons do not share correlations of some sort, as would be the case of truly independent photons.

\begin{figure*}[ht!]
\includegraphics[width=17.6 cm]{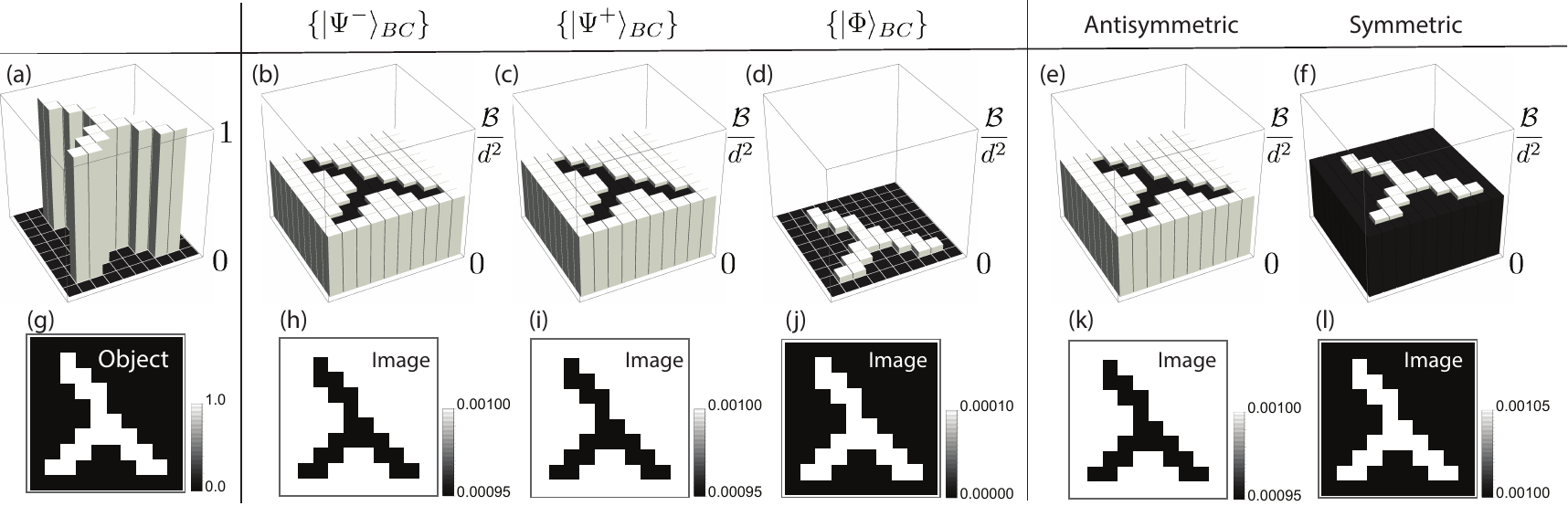}
\caption{\textbf{Predicted ghost images for different projections in path BC.}  (a) Three-dimensional representation of an object in a $d = 100$ dimension space with $\mathcal{B} = 20$. The transmission of each pixel is equal to zero or one. (b)-(f) Three-dimensional representations of the theoretically predicted images.  The height of each pixel represents the probability of the signal. The anti-symmetric image (e) is equal to the image for the projection onto $\{ \ket{\Psi^-}_{BC}\}$ (b); the symmetric image (f) is equal to the sum of the images for the (c) $\{ \ket{\Psi^+}_{BC}\}$ and (d) $\{ \ket{\Phi}_{BC}\}$ projections.  (g)-(l) Two-dimensional representations of the object and images.  In order to highlight the contrast reversal of the image and object, the scales are adjusted for each image.  The heights of the images reflect the true probability of detection, and thus, each image is not normalised, i.e.,~the sum of all pixels in any image does not equal unity. }
\label{fig:simulation}
\end{figure*}

Yet the quantum state of a single photon can be teleported to another, and in an analogous manner, the state of two independent photons can be entangled through a process of entanglement swapping, the latter always resulting in the former \cite{pan1998experimental, jennewein2001experimental}.  In this scenario, which is used in our experiment, two independent pairs of entangled photons are created, with one photon from each pair directed towards a beamsplitter, shown schematically in Fig.~\ref{fig:4way}(b).  On performing a Bell state measurement in paths B and C, the remaining two independent photons in paths A and D become entangled.  This has been demonstrated for both two dimensional and high-dimensional state spaces using both the spin and orbital degrees of freedom of the photon \cite{wang2015quantum, zhang2017simultaneous}, and is anticipated to be a core ingredient of future quantum networks \cite{simon2017towards}.

In this work we demonstrate that the spatial state of non-interacting photons,~i.e.~photons that have never interacted and with no initial position or momentum correlations, can exhibit ghost images that are either contrast-reversed or have the same contrast as the initial object.  We employ entanglement swapping to establish the  correlations needed to reconstruct a ghost image of an object. A notable feature of this work is that the object and reconstructed image can be contrast inverted: bright becomes dark and dark becomes bright.  This is a direct consequence of the correlations that arise after a projection onto the anti-symmetric state during the entanglement swapping process.  We outline the theory behind this phenomenon and find that a projection onto the set of anti-symmetric states results in contrast inversion whereas a projection onto the set  of symmetric states results in conventional ghost imaging, i.e., no contrast inversion.  As a consequence, a projection onto all sets simultaneously, symmetric and anti-symmetric, results in no image at all.   This is due to the inability to establish any spatial correlations between the two independent sources. While image contrast inversion has been observed in a quantum interference experiment \cite{barreto2014}, and simulated by using thermal fermions \cite{liu2016}, this is the first time it has been observed in a ghost imaging setup.  Our work highlights a new form of ghost imaging and paves the way to long distance image transfer across a quantum network.  

\noindent {\bf Theory.}~Consider the four-photon entanglement-swapped ghost imaging setup as shown schematically in Fig.~\ref{fig:4way}(b).  This consists of two independent parametric down-conversion processes that generate two EPR states \cite{howell2004realization}.  Photon pairs A/B, and C/D, are initially entangled in their spatial degrees of freedom, so that the initial four-photon state in the position basis, $\ket{r}$, of dimension $d$ is given by
\bea
\ket{\Psi} = \sum_{i}\frac{1}{\sqrt{d}}\ket{r_i}_A\ket{r_i}_B \otimes \sum_{j}\frac{1}{\sqrt{d}}\ket{r_j}_C\ket{r_j}_D.
\label{eqn:initialstatediscrete}
\eea
\noindent Here the subscripts $i$ and $j$ denote positions at the two crystals, i.e., $r_i$ is the position vector at the SPDC source for photons A/B and similarly for $r_j$.  %Because the crystal planes are imaged to the SLMs in arms A and D these positions also correspond to pixel $i$ and pixel $j$ on SLM A and D, respectively.

We now consider the state of photons in paths B and C, which are incident on a beam splitter BS. A coincidence in the two output paths of the BS projects onto the anti-symmetric states; the absence of a coincidence in the two output paths of the BS projects onto the symmetric states \cite{zhang2016engineering}. In high dimensions $(d>2)$, there are a number of different orthogonal basis in which we can represent the state. The states that we consider are the Bell-like states $\{ \ket{\Psi^{\pm}_{nm}}_{BC} \}$ and $\{ \ket{\Phi_{n}}_{BC} \}$ defined below.  The anti-symmetric states are 
\begin{eqnarray}
\ket{\Psi^-_{nm}}_{BC}  =  \frac{1}{\sqrt{2}}\left[ \ket{r_n}_B\ket{r_m}_C - \ket{r_m}_B\ket{r_n}_C \right], 
\end{eqnarray}
and the symmetric states are
\begin{eqnarray}
 \ket{\Psi^+_{nm}}_{BC}  & = & \frac{1}{\sqrt{2}}\left[ \ket{r_n}_B\ket{r_m}_C + \ket{r_m}_B\ket{r_n}_C \right] \\
 \ket{\Phi_{n}}_{BC} &  =  & \ket{r_n}_B\ket{r_n}_C 
\end{eqnarray}
with $1\leq n < m\leq d$.  We do not consider the Bell-like states $\ket{\Phi^{\pm}_{n,m}}_{BC} = \frac{1}{\sqrt{2}}\left[ \ket{r_n}_B\ket{r_n}_C \pm \ket{r_m}_B\ket{r_m}_C \right]$ as they are not orthogonal to each other for $d > 2$, e.g.~$\langle{\Phi^{+}_{1,2}} \ket{\Phi^{+}_{1,3}} \neq 0$.   The probability to project the photons in path B and C onto an anti-symmetric state is $(d-1)/2d$, whereas the probability to project the photons onto a symmetric state is $(d+1)/2d$ (see supplementary information for more details).

The object corresponds to a transmission mask in path A, where the transmission $O(i)$ at pixel position $i$ is equal to either 1 or 0. The values $\{O(i)\}$ contain all the information about the object. The impact of this mask is to modify the state of photon A such that $ \hat{O}\ket{r_i}_A \rightarrow {O(i)}\ket{r_i}_A$, and the total number of pixels in the mask with transmission equal to 1 is $\sum_i O(i) = \sum_i {O^2(i)} =\mathcal{B}$.

The detection of a photon after the mask in path A combined with the different projections for the two photons in paths B and C can herald different states for photon D.     This is the key to this ghost imaging system: provided that a photon is detected at A, the state of photon D exhibits contrast reversal or not depending on the choice of projection in paths B and C.

We define our ghost images as the set of intensities $\{I_{\Psi^-}(i)\}, \{I_{\Psi^+}(i)\}, \{I_{\Phi}(i)\}, \{I_{AS}(i)\}$ and $\{I_{S}(i)\}$, where each $I(i)$ corresponds to the probability of detecting the light at location $i$ in path D for the different  projections at B and C ($AS = $ anti-symmetric, $S =$ symmetric). Each $I(i)$ can be measured by placing a transmission mask that lets light through at pixel $i$ and blocks all other pixels.  The intensities of the pixels of the images are given by
\begin{eqnarray}
I_{\Psi^-}(i) &=& I_{\Psi^+}(i)  =  \frac{\mathcal{B}-O^2(i)}{2d^2},\\
I_{\Phi}(i)  &=&  \frac{O^2(i)}{d^2},\\
I_{AS}(i)  &=&  \frac{\mathcal{B}-O^2(i)}{2d^2},~{\rm and}\label{e13} \\
I_{S}(i) &=&  \frac{\mathcal{B}+O^2(i)}{2d^2}.\label{e14}
\end{eqnarray}
We see here that the contrast of the image formed from projections onto $\ket{\Psi^-}$ and $\ket{\Psi^+}$ is reversed with respect to that of the object, i.e.,~bright pixels of the object will be measured as dark pixels in the image, and dark pixels of the object will be measured as bright pixels of the image.  Note that a projection onto $\ket{\Psi^-}$  is equivalent to the projection onto the anti-symmetric state. Using Eqs.~\ref{e13} and \ref{e14}, this gives a contrast of the images with respect to the initial object of 
\begin{eqnarray}
\mathcal{C_{AS}} & = & \frac{-1}{\mathcal{B}(d-1)}~{\rm and}~\mathcal{C_{S}} = \frac{1}{\mathcal{B}(d+1)}.\label{contrast}
\end{eqnarray}
These contrasts are of opposite signs, and if the two images are added together, a uniform intensity is a observed with zero contrast.  %This equivalent to performing no projection onto any particular state at BC and then measuring the ghost image.  In this case, as one would expect, no image is observed.   

Figure 2 shows theoretically predicted ghost images for each of the different possible projections for the case of $d = 100$ and $\mathcal{B} = 20$.   The image associated with the anti-symmetric projection ($I_{AS} = I_{\Psi^-}$) exhibits contrast reversal with respect to the object.  Note also that the sum of the images for the anti-symmetric and symmetric images is equal to a flat, constant value that contains no information.  This is as expected from Eq.~(\ref{contrast}), as summing the anti-symmetric and symmetric projections at B and C corresponds to performing no particular projection, leaving no correlation between photons in paths A and D.  Detailed analysis (see supplementary information) shows that the sum of the conditional states $\rho_{AS}$ and $\rho_{S}$ in path D is proportional to the identity matrix and contains no information about the object in path A.

\begin{figure}
\centering
\includegraphics[width=0.95\linewidth]{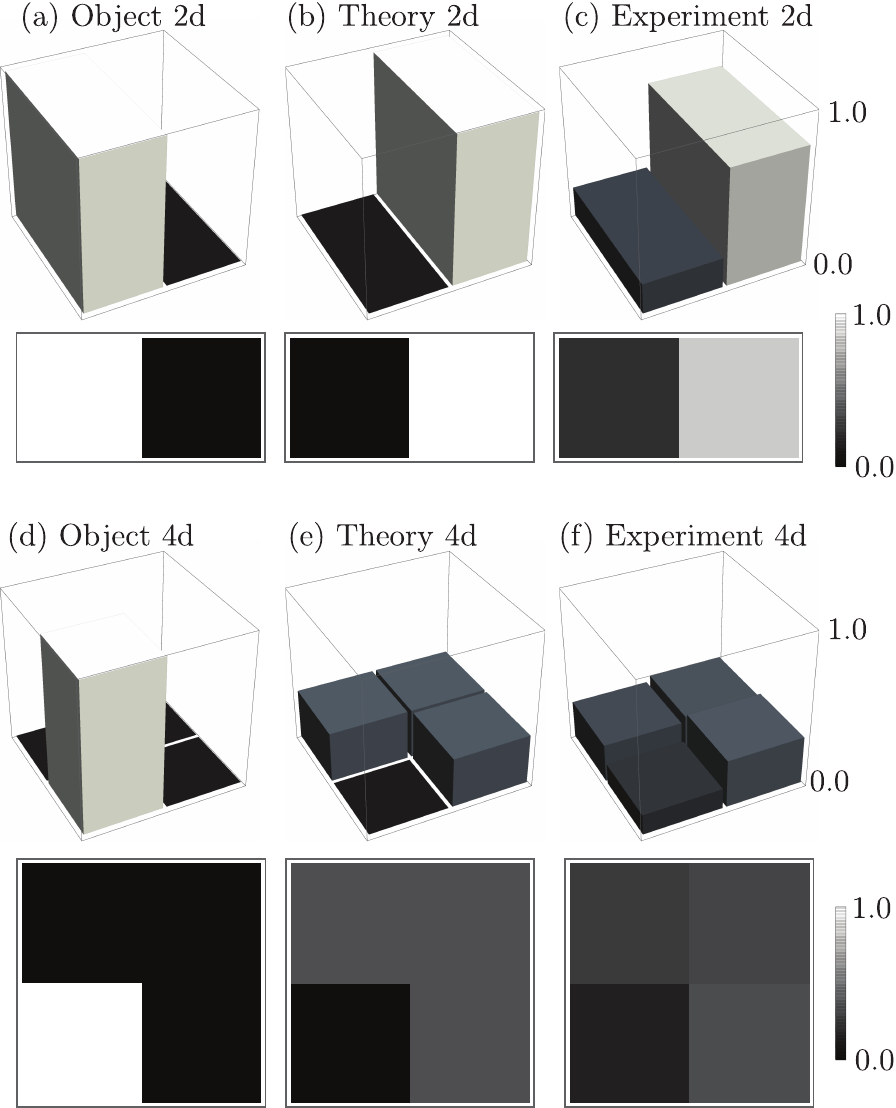}
\caption{\textbf{Contrast-reversed image measured in four-fold coincidence with an anti-symmetric projection.} (a)-(c) $d =2$ data, (d)-(f) $d = 4$ data.  These images are normalised such that the sum of all the pixels in the image is unity.  The total 4-way counts for the $d=2$ image were $\{ 45, 175\}$ counts in 90 mins.  The total 4-way counts for the $d =4$ image were $\{ \{ 168, 191 \}, \{98, 227\} \}$ counts in 800 mins.}
\label{FinalData}
\end{figure}

\noindent {\bf Experimental results.} In our experiment realisation, we establish a four-photon experiment for entanglement swapping, using two BBO (beta barium borate) crystals to produce two pairs of entangled photons.  We choose to project the photons in paths B and C onto the anti-symmetric state and, therefore, measure the images predicted by Eq.~(5) and/or (7).  We examine the behaviour of the system with a simple two-pixel ($d = 2$) image. We relay the crystal planes, which exhibit strong position correlations, onto two spatial light modulators (SLM).  We display an image on SLM A with one half ``on" and one half ``off". On SLM D we display either the same image or the contrast-reversed image. We then measure the number of four-fold coincidences as we vary the path length difference between photons B and C. We see a Hong-Ou-Mandel (HOM) dip in the case where both SLMs have the same pattern on them, while the coincidences stay constant in the case where the SLMs have opposite patterns, see Supplementary Information. This confirms contrast reversal for the anti-symmetric projection.  The corresponding four-photon ghost imaging results are shown in Fig.~\ref{FinalData} (a)-(c), where we compare the object, theoretically predicted image, and the experimentally observed image.  It is clear that the image is contrast reversed with respect to the object.

We also examine the case of a four-pixel image ($d = 4$), with the bottom left quarter of SLM A ``on" and the remainder ``off". On SLM D we scan through one pixel at a time to obtain four measurements, which are then combined to determine the total image observed in photon D. As shown in Fig.~\ref{FinalData} (d)-(f), we obtain a contrast-reversed image in photon D, albeit with a lower contrast than the $d = 2$ case. 

\begin{figure}
\centering
\includegraphics[width=0.95\linewidth]{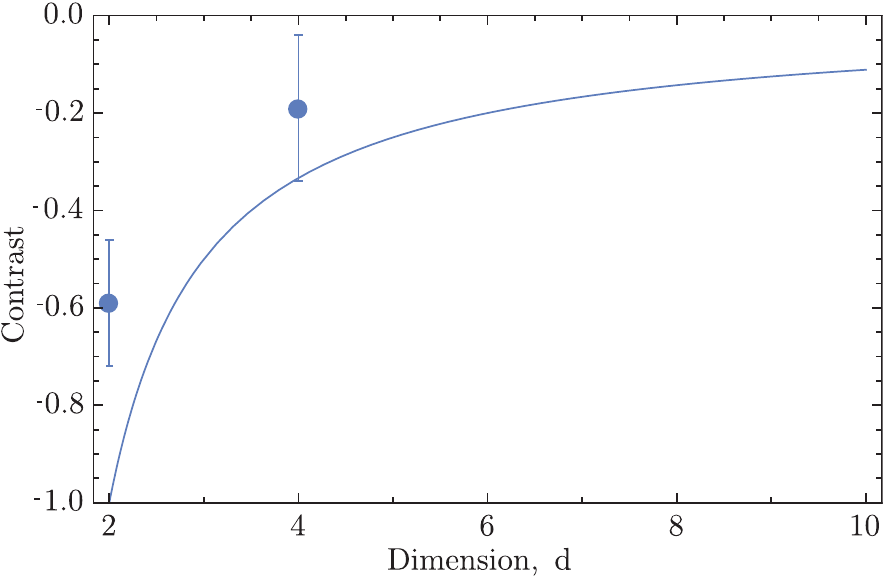}
\caption{\textbf{Contrast as a function of dimension $d$ for anti-symmetric projection.} The data points correspond to the measured contrasts for the data in Fig.~\ref{FinalData}; the solid line is the theoretical prediction according to Eq.~(\ref{contrast}).   The error bars are calculated assuming Poisson noise on the data. }
\label{ContrastData}
\end{figure}

The experimentally recorded images, Fig.~\ref{FinalData} (c) and (f), clearly indicate contrast reversal of the image associated with the anti-symmetric projection in paths B and C.  The measured contrasts for these images are compared to the corresponding predictions in Fig.~\ref{ContrastData}.  We measure a contrast for the $d = 2$ case of $-0.59\pm0.13$ and a contrast for the $d = 4$ case of $-0.19\pm0.15$, which compare favourably to predicted contrast of $-1.0$ and $-0.33$ respectively.  The errors are calculated assuming Poisson statistics on the measured count rates.  Note that the data used in Fig.~\ref{FinalData} is the raw data without any background subtraction applied, and any correction applied to the data will only serve to decrease the measured contrast and therefore increase the agreement between the measured and predicted values. With data processing and subtracting the anticipated 4-way counts based on accidentals, the measured contrasts become $-0.66\pm0.13$ and $-0.26\pm0.14$ for the $d = 2$ and $d = 4$ cases respectively, in good agreement to theory.

%We compare the normalised recovered image \sout{(Fig.~\ref{image}d)} to the theoretically predicted image \av{(see inset of Fig.~\ref{image})} using the mean squared error
%%
%\begin{equation}
%{\rm MSE}=\frac{1}{NM}\sum_{x=0}^N \sum_{y=0}^M \Big[ I(x,y) - P(x,y) \Big]^2,
%\end{equation}
%%
%where $I(x,y)$ is the intensity of the measured image at pixel $\{x,y\}$ and $P(x,y)$ is the intensity of the predicted image at pixel $\{x,y\}$. For our particular predicted image, the MSE can range from zero for a measured image identical to the prediction and $1/3$ for a measured image with no similarity. For our measured image, we obtain ${\rm MSE}=0.0078$.
%

x

\noindent {\bf Discussion.} Quantum teleportation is defined as teleporting an exact state to a new location.  An implementation of quantum teleportation only requires three photons, but it is conventional to start with a four-photon system such as the one outlined in our experiment.  Here, we use arm A to herald an object in arm B that we wish to teleported to arm D.  In two dimensions, we do indeed teleport an exact copy of the state to a new location.  When we consider the high-dimensional case, the photons in A and D are left in the mixed state of all possible two-photon two-dimensional anti-symmetric states \cite{zhang2017simultaneous}, so that knowledge of the state of the photon in A now no longer gives precise knowledge of the state of the photon in D - it only gives partial knowledge.  This can be seen in the four-dimensional example: while the measured image is the contrast reversed version of the object, it is not an exact copy.  Perfect image teleportation in any dimension could be achieved through mimicking standard ghost imaging by applying an accumulative approach,  reconstructing the object through successive two-dimensional projections; by using multiple photons in an extended version of our experimental set-up (e.g., 5 for $d=3$, 6 for $d=4$ and so on) which has been theorised to allow high-dimensional teleportation \cite{goyal2014qudit}; or by implementing a state selective projection in arms B and C \cite{zhang2017simultaneous}.     

While teleporting images through an entanglement swapping process is a necessary step for transmitting information across a quantum network, such four-photon experiments present some challenges.   Our experiment suffers from low count rates due in large part to measuring a single pixel at a time. In the case of the two-pixel image, we can at maximum detect $p=50\%$ of the total single counts at any one time; in the four-pixel case, that number is 25\%. As this reduction occurs in both arms A and D, and because the detection of these modes also reduces the counts in arms B and C by the same amount, we see a $p^4$ reduction in the four-fold coincidences, which in the four-pixel case is a factor of 0.004. For this reason, increasing the resolution of the image makes it prohibitively time-consuming to obtain sufficient counts.  Further, the theory for our experiment predicts a decreasing contrast with dimension, which together with the aforementioned limitations makes accessing larger dimensions difficult.  However we predict that these technical challenges can be overcome, and that the observed contrast decrease with dimension can be ameliorated as outlined above.

To conclude, this experiment constitutes the first implementation of ghost imaging using independent photons, and the first observed contrast reversal in ghost imaging.  Our image is teleported from arm A to arm D exactly in two dimensions but it is a more complex scenario in higher dimensions.  This results in reduced contrast images for large dimensions, which we predict can be ameliorated by judicious control of the projections for photons B and C.  This work represents an important step to realise image transfer across a quantum network.  %While we do use the same laser to produce both pairs of entangled photons, the result would be identical using two different lasers.

\section{Acknowledgements}

We thank Prof. Daniele Faccio and Dr. Yingwen Zhang for helpful discussions.  AV was supported by the Claude Leon Foundation; NB was supported by the DST-CSIR; AF was supported by NRF; JL was supported by the Engineering and Physical Sciences Research Council through the Quantum Hub in Quantum Enhanced Imaging (EP/M01326X/1).

%\bibliographystyle{apsrev4-1}
%\bibliography{References_File}


\newpage

\end{document}